\documentclass[aps, twocolumn]{revtex4-1}

\usepackage{graphicx}
\usepackage{amsmath}
\usepackage{dsfont}
\usepackage{amssymb}
\usepackage{physics}
\usepackage{hyperref}
\usepackage{ulem}
\hypersetup{colorlinks=true,
	    final=true,
	    linkcolor=blue,
	    citecolor=blue,
	    filecolor=blue,
	    urlcolor=blue,}
\usepackage{amssymb}
\usepackage{subfigure}

\newcommand{\rta}{\rightarrow}

\newcommand{\ep}{\epsilon}

\newcommand{\p}{\prime}
\newcommand{\om}{\omega}
\newcommand{\ra}{\rangle}
\newcommand{\la}{\langle}

\newcommand{\beq}{\begin{equation}}
\newcommand{\eeq}{\end{equation}}

\newcommand{\mathq}{\mathbf{q}}
\newcommand{\mathk}{\mathbf{k}}

\begin{document}

\title{On the T-linear resistivity of cuprates: theory}
\author{Charu Dhiman$^1$, Raman Sharma$^1$, and Navinder Singh$^2$,}
\email{navinder.phy@gmail.com; Phone: +91 9662680605}
\affiliation{$^1$ Department of Physics, HP university, Shimla, HP, and \\ $^2$Theoretical Physics Division, Physical Research Laboratory (PRL), Ahmedabad, India. PIN: 380009.}

\begin{abstract}
By partitioning the electronic system of the optimally doped cuprates in two electronic components: (1) mobile electrons on oxygen sub-lattice; and (2) localized spins on copper sub-lattice, and considering the scattering of mobile electrons (on oxygen sub-lattice) via generation of paramagnons in the localized sub-system (copper spins), we ask what should be the electron-paramagnon coupling matrix element $M_q$ so that T-linear resistivity results. This 'reverse engineering approach' leads to $|M_q|^2 \sim \frac{1}{q^2+\xi(T)^{-2}}$. We comment how can such exotic coupling emerge in 2D systems where short range magnetic fluctuations resides. In other words, the role of quantum criticality is found to be crucial. And the T-linear behaviour of resistivity demands that the magnetic correlation length scales as $\xi(T)\propto\frac{1}{T}$, which seems to be a reasonable assumption in the quantum critical regime of cuprates (that is, near optimal doping where T-linear resistivity is observed).
\end{abstract}

\maketitle


Electronic transport is a fascinating subject\cite{ziman,toda,ted,singh,kubo,ziman2}. It is commonly said that resistivity is one of the first properties to be measured in a material, yet often one of the last to be fully understood. One of the basic models to analyze resistivity is the Drude model\cite{singh,ashcroft, navkomal, navdrude}:

\beq
\sigma(\om) = \frac{ne^2}{m}\frac{1}{i\om +\gamma}.
\eeq

Here, $\gamma$ is the Drude scattering rate, $n$ is the number of electrons per unit volume.  This can be derived in various ways\cite{navkomal,singh}, but it's derivation using Langevin equation is quite insightful\cite{navkomal}:

\beq
m \dot{u}(t) = \underbrace{- m\gamma u(t) +f(t)}_{systematic ~~part} + \underbrace{R(t)}_{random~~part.}
\eeq     

Electron is treated as a 'Brownian particle' on which systematic  ($-m\gamma u(t) +f(t)$) and random force ($R(t)$) acts.  $f(t)$ is the force coming from applied electric field. It is well known that Langevin equation suffers from a defect: The Langevin equation violates the requirement of stationarity in an equilibrium setting\cite{singh,navkomal}.  Hence, the Drude formula (derived from it) is bound to fail in a general setting\cite{navkomal}. The more accurate approach is to appeal to the generalized Langevin equation\cite{singh}:

\beq
m \dot{u}(t) = \underbrace{- m \int_{-\infty}^t dt^\p M(t-t^\p) u(t^\p) +f(t)}_{systematic~part~with~memory} + \underbrace{R(t)}_{random~~part}.
\eeq

Here $M(t)$ is time dependent friction coefficient (the memory function). The solution of the generalized Langevin equation (which does not suffer from stationarity problem) for conductivity can be obtained easily with Fourier transforms\cite{navkomal}:

\beq
\sigma(\om) = \frac{ne^2}{m}\frac{1}{i\om +M(\om)}.
\eeq
Where $M(\om) = \int_0^\infty dt M(t) e^{-i\om t -0^+t}$ is the Fourier-Laplace transform of $M(t)$.  This is called the generalized Drude formula (GDF)\cite{singh,gw}. In the case of no memory $M(t) = \gamma \delta(t)$, the GDF goes back to simple Drude formula.

GDF can also be derived from the Kubo formula for conductivity
\beq
\sigma_{ab}(\om) = \int_0^\infty dt e^{i\om t}\int_0^\beta d\lambda \la\ I_a(-i\hbar\lambda) I_b(t)\ra
\eeq
using the Projection-operator techniques\cite{singh}. In the above equation $\beta =\frac{1}{k_B T}$ and $I$ is the current operator. On the other hand, it turns out that the Fourier-Laplace transform of the current-current correlation function
\beq
\chi(\om) = i\int_0^\infty dt e^{i\om t-0^+t}\la[I(t),I(0)]\ra
\eeq
can be expressed in terms of the memory function\cite{singh,gw}: 
\beq
\chi(\om) = \chi(0)\frac{M(\om)}{\om + M(\om)}
\eeq

Therefore, the computation of the Fourier-Laplace transform of the current-current correlation function leads to the expression of the memory function,  thus conductivity (equation 4) can be computed. The imaginary part of the memory function gives the generalized Drude scattering rate ($M''(\om,T) =\frac{1}{\tau(\om,T)}$).  And DC resistivity is given as $\rho(T) = \frac{m^*}{ne^2}\frac{1}{\tau(T)} = \frac{m^*}{ne^2} M''(T)$. $M''(T)$ is the D.C. limit of $M''(\om,T)$.

In 1972, Goetze and Woelfle\cite{gw,singh}, came up with a perturbative expansion of the memory function. It turns out that

\beq
M(z) \simeq \frac{1}{z\chi(0)}[\la\la C;C\ra\ra_0 -\la\la C;C\ra\ra_z].
\eeq

Here the double-angle-brackets is the notation for: $\la\la A;B\ra\ra_z = i\int_0^\infty dt e^{i zt} \la[A(t),B(0)]\ra$.  $z= \om +i0^+$ is the complex frequency. And $C= [I, H_{int}]$ is the commutator, in which the current operator is given by $I =\sum_\mathk v_x(\mathk)c_\mathk^\dagger c_\mathk$ and $H_{int}$ is the interaction part of the Hamiltonian (for details of the formalism, refer to\cite{singh,gw}).

With this background to the memory function technique, let us review the electronic situation in the $CuO_2$ planes of cuprates near the optimal doping\cite{navcup,emery}. The partitioning of the electronic system of the optimally doped cuprates in two electronic components: (1) mobile electrons on oxygen sub-lattice; and (2) localized spins on copper sub-lattice is supported by extensive experimental evidence\cite{navcup,emery,sun}. Consider the scattering of mobile electrons (on oxygen sub-lattice) via generation of paramagnons in the localized sub-system (copper spins), near optimally doped regime.  Write the Hamiltonian in the following way: 

\begin{eqnarray}
 H = H_{ele} &+& H_{paramagnon} + H_{ele-paramagnon} .\\
 H_{ele} &=& \sum_\mathk \ep_\mathk c_\mathk^\dagger c_\mathk.\\
 H_{paramagnon} &=& \sum_\mathq \omega_\mathq \left( b_\mathq^\dagger b_\mathq + \frac{1}{2} \right). \\
 H_{ele-paramagnon} &=& \sum_{\mathk,\mathk'} \left( M(\mathk - \mathk') c_\mathk^\dagger c_{\mathk'} b_{\mathk - \mathk'} + \text{h.c.} \right).\nonumber\\
\end{eqnarray}
Notation is standard (we set $\hbar=1$). We use the quadratic dispersion for the mobile component. $M(\mathq)$ is the matrix element of the electron-paramagnon coupling. The form of which we will specify at the end of the calculation (there we will look into the quantum critical aspects near optimal doping). 

The expression for $C$ can be computed as
\begin{equation}
C= [I_x,H]=[I_x, H_{ele} + H_{paramagnon} +  H_{ele-paramagnon}]
\end{equation}

The non-interacting parts of Hamiltonian commute with current operator, thus the expression of $C$ takes the form

\begin{equation}
 C=\sum_{\mathk,\mathk'} [v_x(\mathk) - v_x(\mathk')]M(\mathk-\mathk') c_\mathk^\dagger c_{\mathk'} b_{\mathk - \mathk'} -h.c.
 \end{equation}

Substitute this expression of $C$ into equation (8). After a lengthy but straightforward calculation\cite{singh}, the expression for the imaginary part of the memory function (equation 8) takes the form:

\begin{widetext}
\beq
M''(\omega) = A \sum_{\mathk,\mathk'} \left| M(\mathk - \mathk') \right|^2 \left| \mathk - \mathk' \right|^2 (1 - f_k) f_{k'} n_{\mathk-\mathk'} \\
\times \left[ \left( \frac{e^{\beta \omega} - 1}{\omega} \right) \delta(\epsilon_k - \epsilon_{k'} - \omega + \omega_{\mathk-\mathk'}) + \text{(term} ~\om\rta -\om) \right]
\eeq
\end{widetext}

All the constants we have collected in coefficient $A$. To simplify the calculation insert $\int_0^\infty dq \delta(q-|\mathk-\mathk'|) =1$, and convert summations over $\mathk,~\mathk'$ into integrals. Fix $\mathk$ along the z-axis an assume $\mathk'$ subtends an angle $\theta$ with it. First integrate over $\mathk'$ and then over $\mathk$. The resulting expression takes the form:

\begin{widetext}
\begin{eqnarray}
M''(\omega) &=& A\int_0^\infty dq \sum_\mathk \int dk' k' \int_0^{2\pi} d\theta\delta(q-|\mathk-\mathk'|) |M(q)|^2 q^2  n_q   (1 - f_k) f_{k'} \nonumber\\
&\times& \left[ \left( \frac{e^{\beta \omega} - 1}{\omega} \right) \delta(\epsilon_k - \epsilon_{k'} - \omega + \omega_q) + \text{(term} ~\om\rta -\om) \right].
\end{eqnarray}
\end{widetext}

If we consider physically relevant case: $\ep_F>>k_B T$, that is, Fermi energy much much greater than thermal energy scale, then the product of Fermi factors $(1-f_k)f_k'$ acts roughly like a delta function situated at the Fermi energy. With this information, the integral over the delta function can be quickly simplified as: 

\beq
\int_0^{2\pi} d\theta\delta(q-|\mathk-\mathk'|) \simeq \frac{1}{k_F} \int_0^\infty d\theta \delta(\frac{q}{k_F} - \sqrt{2(1-\cos(\theta))})
\eeq
Which can be simplified by finding the roots of the equation $\theta =\cos^{-1}(1-\frac{q^2}{2k_F^2})$ and using the expression $\delta(f(x)) = \sum_n \frac{\delta(x-a_n)}{|f'(a_n)|^2}$, where $a_n$ are the roots of $f(x)$. The result is $1/k_F$.  In 2D (relevant to $CuO_2$ planes) the electronic density of states for parabolic band is constant $\rho(\ep) =\frac{dn}{d\ep} =Const.$. Converting integrals over $k$ into integrals over energy $\int d^2\mathk \rta constant\int d\ep,~~\int k'dk' \rta constant\int d\ep'$, and the integral over $\ep'$ can be strait-forwardly done using the properties of the delta function in the above equation. The integral over $d\ep$ can be done using standard tables $\int_{-\infty}^{\infty} dx \frac{e^x}{e^x+1} \frac{1}{e^{x+a}+1} = \frac{a}{e^a-1}$. We finally obtain:

\begin{widetext}
\beq
M''(\omega) = A\int_0^{q_{cut}} dq |M(q)|^2 q^2 n_q \left[\frac{e^{\beta \om} - 1}{\om}\frac{\om-\om_q}{e^{\beta(\om-\om_q)}-1} + \frac{e^{-\beta \om} - 1}{\om}\frac{\om+\om_q}{e^{-\beta(\om+\om_q)}-1}  \right].
\eeq
\end{widetext}

Here, we introduced the legitimate upper cut-off on the paramagnon wave-vector: $q_{cut} \simeq \frac{\pi}{a}$, where $a$ is the lattice constant. On taking the DC limit $\om\rta 0$, and after some calculation, we get

\beq
\frac{1}{\tau(T)} = M''(T) = A \beta \int_0^{q_{cut}} dq  |M(q)|^2 q^2 \frac{\om_q}{\cosh(\beta\om_q)-1}.
\eeq

This is the main result of our work. To investigate how does DC resistivity $\rho(T) = \frac{m^*}{n e^2}\frac{1}{\tau(T)}\propto M''(T)$ scale with temperature we consider two special cases of interest:

\subsection*{Case A:  Regular electron-paramagnon coupling matrix element in the paramagnetic state of a 2D magnetic metal}

In the case of a magnetic metal in the paramagnetic state (away from the any critical point) the electron-paramagnon coupling matrix element can be taken as a constant ($\mathq$ independent)\cite{moriya}. In this case, the above equation (eqn(19)) simplifies to

\beq
M''(T) = A M^2 \beta \int_0^{q_{cut}} dq q^2 \frac{\om_q}{\cosh(\beta\om_q)-1}.
\eeq

In the AFM case, the paramagnon dispersion is given by $\hbar\om_{q} = \hbar D_1 q$, where $D_1$ spin stiffness constant.  In the FM case: $\hbar\om_q = \hbar D_2 q^2$. In the high temperature limit $k_B T>>\hbar \om_{cut} = \hbar D_1 q_{cut}$; and $k_B T>>\hbar \om_{cut} = \hbar D_2 q_{cut}^2$, it is easy to see (by expanding the $\cosh(x)$) that the above expression scales linearly with temperature:

\beq
M''(T) \propto T.
\eeq

In the low temperature limit, the scaling behaviour of the imaginary part of the memory function depends the dispersion used. For AFM case, we get (by changing variables: $x = \beta D_1q$):

\beq
M''(T) \propto (k_B T)^3\int_0^{\frac{D_1q_{cut}}{k_B T}}dx \frac{x^3}{\cosh(x)-1}\propto T^3.
\eeq

In the considered low temperature limit $k_B T<< D_1 q_{cut}$, the upper limit of the integral in the above expression can be considered $\infty$. Thus in the AFM case, the resistivity of a 2D magnetic metal in the paramagnetic state will scales as $T^3$. It can be easily checked that in the FM case, in the low temperature limit, resistivity scales as $T^{3/2}$.

\subsection*{Case B: Singular electron-paramagnon coupling matrix element in a metal near a magnetic quantum criticality}

Much more interesting case emerges if the electron-paramagnon matrix element has non-trivial dependence on wave vector $q$ such as when the system is near a quantum critical point. It turns out that\cite{chubukov1,chubukov2}, for systems near magnetic quantum critical points, the electron-paramagnon matrix element can be taken as

\beq
|M(q)|^2 \sim \chi(q) \sim \frac{1}{q^2 + \xi(T)^{-2}}
\eeq

Here $\xi(T)$ is the magnetic correlation length which diverges at the QCP. We are in the finite temperature regime, and use a value which is known from neutron magnetic scattering experiments (typically of a few lattice constants at the optimal doping at room temperature). From this information we deduced the coefficient $D$ in the expression: $\xi(T) = \frac{D}{k_B T}$ . With this matrix element, equation (19) takes the form:

\beq
M''(T) = A \beta \int_0^{q_{cut}} dq  \left(\frac{q^2}{q^2 + \xi(T)^{-2}}\right) \frac{\om_q}{\cosh(\beta\om_q)-1}.
\eeq

This is our second main result. We focus on the AFM regime relevant to cuprates.

Question is what will happen in the low temperature regime, when the correlation length becomes very large? In the low temperature regime $k_B T<< D_1 q_{cut}$, substitute $x=\beta D_1 q$, the integral takes the form:

\beq
M''(T) \propto k_B T \int_0^{\frac{D_1q_{cut}}{k_BT}} dx \frac{x^2}{x^2 + \left(\frac{D_1/\xi(T)}{k_B T}\right)^2}\left(\frac{x}{\cosh(x)-1}\right)
\eeq

Our next crucial assumption is that the correlation length is inversely proportional to temperature\cite{com1}. We set $\xi(T) = \frac{D}{k_B T}$. With this, the above expression takes the form:

\beq
M''(T) \propto k_B T \int_0^\infty dx \frac{x^2}{x^2 + (\frac{D_1}{D})^2}\left(\frac{x}{\cosh(x)-1}\right).
\eeq

The upper limit is extend to $\infty$ as $k_B T<< D_1 q_{cut}$. The integral is convergent and we get perfectly $T-$linear behaviour of the scattering rate, thus resistivity, in the low temperature limit. This is our min result. The scattering rate can be written as: $M''(T) \propto g(\frac{D_1}{D}) k_B T$. Here, $g(\frac{D_1}{D})$ is a function which depends on material specifics (the value of spin stiffness etc).

In the high temperature limit ($k_B T>> D_1 q_{cut}$) , we can expand $\cosh(x)$ in equation (25), as $x =\beta D_1q<<1$, and we get:

\beq
M''(T)\propto k_BT \ln(1+ \left(\frac{Dq_{cut}}{k_B T}\right)^2).
\eeq

The temperature dependence now is logarithmically suppressed in the limit: $Dq_{cut}>>k_BT$. In the opposite limit, we observe $M''(T)\propto\frac{1}{T}$. However, we argue that the physically relevant regime for optimally doped cuprates corresponds to: $k_B T<<D_1 q_{cut}~\sim~D q_{cut}$, where $D_1$ and $D$ are of similar order. This energy scale is much higher. For example, in the undoped $La_2CuO_4$, the stiffness constant (spin wave velocity) is $D_1=850~meV.~\AA$ ($\om_q = D_1 q$)\cite{col}. The energy scale $\om_{cut} =D_1 q_{cut}$ is of the order of $0.7~eV$ ($\sim8000~K$!) for $q_{cut} = \frac{\pi}{a}, ~~a=3.8~\AA$. Also, if we consider 2D AFM Heisenberg model, the spin stiffness (spin wave velocity) scales as $D_1=2\sqrt{2}J S a$, where $J$ is the superexchange interaction, $S$ is the magnitude of spin, and $a$ is the lattice constant\cite{aur}. From the theoretical model, the value of $D_1q_{cut}$ turns out to be $\sim 5000~K$ for typical value of $J=0.13~eV$. This is also much greater than the temperature scale on which DC resistivity is measured (from $milli~Kelvin$ to $\sim 1000~K$).

In the doped case (especially optimally doped cuprates) spin waves are diffusive (paramagnons are damped and decay). However, an effective spin stiffness (much less in magnitude) can be defined\cite{dean}.  An effective value of $D_1$ for doped YBCO is $\sim200~meV. \AA$\cite{dean}. The upper cut-off energy scale now is: $D_1 q_{cut}\sim 160~meV \sim2000~K$ which is still higher than the measured temperatures.

The full plot of the resistivity for physically relevant parameters is shown in figure (1).  We take the value of $D_1 = 200~meV. \AA$\cite{dean}. The value of $D$ is taken to be $500~meV.\AA$ which corresponds to magnetic correlation length of about five lattice constants at optimal doping at room temperature.

\begin{figure}[!tbp]
  \centering
  \begin{minipage}[b]{0.5\textwidth}
    \includegraphics[width=\textwidth]{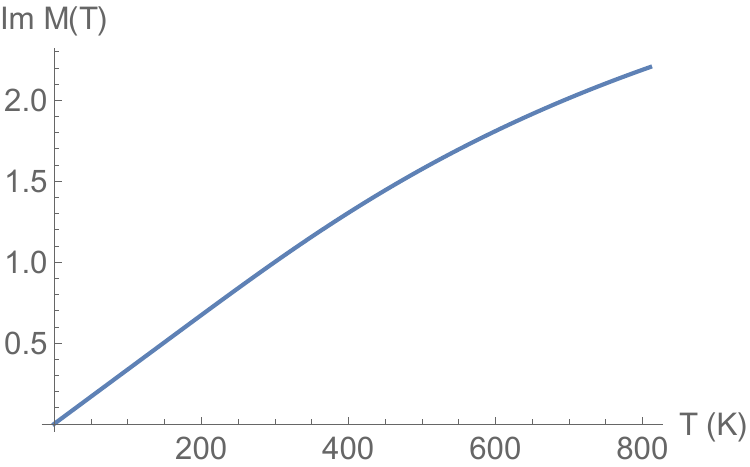}
    \caption{The scattering rate for: upper cut-off for the paramagnon energy scale ($\hbar\om_{cut} = D_1 q_{cut} \simeq 160~meV$)\cite{dean}.}
  \end{minipage}
  \hfill
  \begin{minipage}[b]{0.5\textwidth}
    \includegraphics[width=\textwidth]{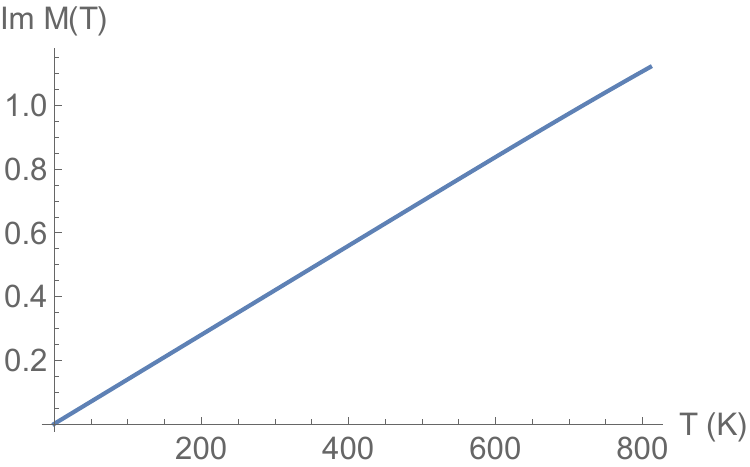}
    \caption{The scattering rate for: upper cut-off for the paramagnon energy scale ($\hbar\om_{cut} = D_1 q_{cut} \simeq 300~meV$)\cite{tacon}.}
  \end{minipage}
\end{figure}

In YBCO, using RIXS, dispersive paramagnon energy scale as large as $300~ meV$ has been reported\cite{tacon}. If we use this value as the upper cut-off (with $q_{cut} =\frac{\pi}{a}$) we get $D_1 = 400~meV.\AA$, and we obtain linear behaviour even upto 800 K! (figure 2).

In conclusion, we have the following points to make:
\begin{enumerate}
\item T-linear resistivity of cuprates originates from the scattering of mobile carriers on oxygen sub-lattice via generation of paramagnons in the fluctuating spins of the copper sub-lattice.
\item $T\rta0$ perfect T-linear behaviour is attributed to the quantum critical point deep under the superconducting dome where magnetic correlation diverges. Our theory demands that $\xi(T)\propto\frac{1}{T}$.
\item There seems to be no saturation in the high temperature limit (even upto 800 K) of the resistivity. This we attribute to very high energy scale of the paramagnon cut-off energy (the analogue of the Debye scale in the case of electron-phonon scattering). This upper cut-off can be as large as $300~meV\sim3000~K!$
\end{enumerate}
We think that this should constitute a reasonable case for the explanation of the $T-linear$ behaviour of DC resistivity. Further open problem (which can be solved within this formalism) is that of the optical conductivity $\sigma(\om)\sim\frac{1}{\om}$, which will be taken up in a future investigation.

\vspace{1cm}

Authors would like to thank A.-M. S. Tremblay for useful correspondence and important comments.


\begin{thebibliography}{201}
\bibitem{com1} It is observed that for the 2D Hubbard model on a square lattice, using Two-Particle Self-Consistent (TPSC) theory, the scaling behaviour of the correlation length is $\xi(T)\sim \frac{1}{T}$, which corresponds to $z=1$ scaling regime near the SDW instability\cite{dom}.
\end{thebibliography}
\end{document}